\documentclass{article}

\usepackage{PRIMEarxiv}

\usepackage[utf8]{inputenc} 
\usepackage[T1]{fontenc}    
\usepackage{hyperref}       
\usepackage{url}            
\usepackage{booktabs}       
\usepackage{amsfonts}       
\usepackage{nicefrac}       
\usepackage{microtype}      
\usepackage{lipsum}
\usepackage{fancyhdr}       
\usepackage{graphicx}       
\graphicspath{{media/}}     
\usepackage{amsmath}
\usepackage{algorithm2e}
\usepackage{algpseudocode}
\usepackage[subtle]{savetrees}
\usepackage{subfig}
\usepackage{float}
\usepackage{overpic}
\usepackage{tabularx} 

\DeclareMathOperator*{\argmin}{arg\,min}
\RestyleAlgo{ruled}
\title{Deconvolving Complex Neuronal Networks into Interpretable Task-Specific Connectomes
}

\author{
  Yifan Wang,  Ananth Grama\\
  Department of Computer Science,\\
  Purdue University,\\
  West Lafayette, IN, USA \\
  \texttt{\{wang5617, ayg\}@purdue.edu} \\
   \And
  Vikram Ravindra \\
  Department of Computer Science,\\
  University of Cincinnati,\\
  Cincinnati, OH, USA \\
  \texttt{vikram.ravindra@uc.edu} \\
}

\begin{document}
\maketitle


\begin{abstract}
Task-specific functional MRI (fMRI) images provide excellent modalities for studying the neuronal basis of cognitive processes. We use fMRI data to formulate and solve the problem of deconvolving task-specific aggregate neuronal networks into a set of basic building blocks called canonical networks, to use these networks for functional characterization, and to characterize the physiological basis of these responses by mapping them to regions of the brain. Our results show excellent task-specificity of canonical networks, i.e., the expression of a small number of canonical networks can be used to accurately predict tasks; generalizability across cohorts, i.e., canonical networks are conserved across diverse populations, studies, and acquisition protocols; and that canonical networks have strong anatomical and physiological basis. From a methods perspective, the problem of identifying these canonical networks poses challenges rooted in the high dimensionality, small sample size, acquisition variability, and noise. Our deconvolution technique is based on non-negative matrix factorization (NMF) that identifies canonical networks as factors of a suitably constructed matrix. We demonstrate that our method scales to large datasets, yields stable and accurate factors, and is robust to noise.
\end{abstract}
\keywords{deconvolving connectomes, matrix factorization, dimension reduction, functional MRI.}

\section{Introduction}
Connectomic studies use functional brain images of subjects performing tasks to understand complex cognitive processes. Functional Magnetic Resonance Imaging (fMRI) is a common imaging modality used to analyze the underlying natural processes in healthy individuals and the dysregulation of such processes due to disease and/or injury. Functional networks derived from fMRIs typically superpose many neurophysiological responses elicited by stimuli. Identifying these responses and separating associated functional networks into their basic building blocks is essential to understanding the shared, and unique aspects of neuronal responses across heterogeneous populations performing different tasks. Ideally, these canonical networks must be grounded in neurophysiology, identifying coherent modules of neural responses that are interpretable by neuroscientists and other domain experts.

The method of choice for connectomic analysis is Independent Component Analysis (ICA) \cite{makeig95,hyvarinen00}, which is used on individual fMRIs to spatially localize regions of interest. Group-ICA \cite{esposito05,du13,lin10} combines fMRIs across individuals to model shared regions of interest. Other ML-based interpretable methods have been proposed in the recent past \cite{grosenick08, jain20, li21, kan22}. However, these methods are limited in their ability to handle large datasets with diverse subjects (young v/s old, healthy v/s diseased) performing a variety of cognitive tasks. Moreover, most of these learned representations lack inherent interpretability or transparency to yield insights into brain elemental networks associated with different tasks. Large-scale efforts, such as the Human Connectome Project \cite{van13}, Cambridge Centre for Ageing and Neuroscience (CamCAN)~\cite{taylor17}, and Alzheimer's Disease Neuroimaging Initiative (ADNI) \cite{jack08} have each collected and curated neuroimages from cohorts of several hundred subjects. 
Each of these datasets includes subjects of different ages, stages of neuroplasticity, and degeneration. These datesets provide us with excellent opportunities for developing methods for identifying canonical functional networks, their compositions, and use in diffrent applications.

In this paper, we propose a novel framework for deconvolving networks derived from fMRIs of subjects performing different tasks into a small set of canonical networks that serve as building blocks that are: (i) shared across large cohorts; (ii) can be composed into task-specific networks; and (iii) are predictive of tasks and efficacy. We call these networks \emph{canonical task connectomes}. Our framework also characterizes the extent of expression of these networks for each task, along with its neurophysiological basis.

Our approach first combines individual functional networks into a population-level matrix $\mathbf{X}$. We then deconvolve this matrix into its factors $\mathbf{W}$ and $\mathbf{H}$ such that each column $\mathbf{W}_{(\ast,i)}$ corresponds to a canonical task connectome, and the corresponding row $\mathbf{H}_{(j,\ast)}$ characterizes the extent to which the canonical network is expressed in every subject. However, since individual samples (fMRIs) correspond to subjects performing different tasks, the latent canonical representations must encode this important information. We accomplish this by formulating and solving a suitable matrix factorization problem.

We present our main experimental results on 1000 subjects from HCP at rest and for six tasks (Language, Emotional Processing, Gambling, Motor, Relational Processing, and Social Processing). Our results show that: 
\begin{itemize}
    \item \emph{Canonical task connectomes have high task-specificity}. We show that our approach constructs networks that uniquely characterize different tasks and are therefore excellent markers of tasks. 
    \item \emph{Canonical task connectomes are generalizable across cohorts.} We show that canonical representations obtained on a suitably constructed train set can accurately predict tasks being performed by the test set. 
    \item \emph{Canonical task connectomes identify common neural processes.} We show that our approach finds canonical functional networks that are shared across tasks. This enables novel interpretations of processes and responses associated with different task stimuli. 
    \item \emph{Canonical task connectomes have a strong physiological basis.} We show that the canonical connectomes can be mapped to regions of the brain to identify physiological underpinnings of tasks, that are in strong agreement with literature in neurosciences.
\end{itemize}

The rest of the paper is organized as follows: in Section \ref{subsec:nmf}, we summarize non-negative matrix factorization and its use in our context. In Section \ref{subsec:framework}, we provide details for our proposed framework. Then, we describe the HCP dataset and the preprocessing pipeline in Section \ref{subsec:data}. This is followed by comprehensive experimental validation in section \ref{sec:result}, where we demonstrate the interpretability and generalizability of our proposed approach. Finally, we conclude with related work in Section \ref{sec:related} and summarize our contributions and avenues for future work in Section \ref{sec:conclusion}.

\section{Methods and Materials}
\label{sec:method}
We describe our formulation and solution to the problem of identifying interpretable task-specific brain networks, called ``connectomes'' from neuroimaging datasets of subjects performing a variety of cognitive tasks. Connectomes are networks in which regions of the brain correspond to nodes and correlated activity quantifies the strength of edges across corresponding nodes (regions). We describe, in more detail, the process of construction of connectomes in Section~\ref{subsec:data}.

We hypothesize and validate that neuronal activity observed during a task is composed of a small set of canonical patterns or motifs that {\em recur across subjects}. Correspondingly, the overall observed connectome is a superposition of these motifs that we call canonical task connectomes. The goal of our formulation and methods is to demonstrate the existence and applications of such canonical task connectomes. 

We abstract the connectome into a $region \times region$ similarity matrix. Applying non-negative matrix factorization to connectome data from a large group of subjects, we factor the observed composite connectome into canonical task connectomes. These factors, extracted by non-negative matrix factorization (NMF), provide consistent and strong associations with specific cognitive tasks. This provides strong evidence for our framework of canonical task connectomes, while yielding physiologically interpretable results.

\subsection{Non-negative Matrix Factorization (NMF)}
\label{subsec:nmf}
Let $\mathbf{X} \in \mathbb{R}^{p \times n}_{\geq 0}$ denote the data matrix. NMF approximately factorizes the data matrix $\mathbf{X}$ into two non-negative matrices $\mathbf{W}$ and $\mathbf{H}$ by optimizing:

\begin{equation}
\argmin_{\mathbf{W},\mathbf{H}\geq0} ||\mathbf{X}-\mathbf{W} \mathbf{H}||_F^2 
\end{equation}

Where, $\mathbf{W} \in \mathbb{R}^{p \times r}_{\geq 0}$ is the (non-negative) ``basis matrix'' which is a low-rank, latent description of the columns of $\mathbf{X}$, 
$\mathbf{H} \in \mathbb{R}^{r \times n}_{\geq 0}$ is the (non-negative) ``encoding matrix'' matrix of coefficients that provides the weights to each of the latent factors required to explain each data-point. Here, $r$ is the rank, representing the number of latent factors used to approximate the original data matrix $\mathbf{X}$. 

The non-negativity constraints in NMF lead to a parts-based representation of the data unlike ICA, which allows positive and negative values. This enforces sparsity and non-subtractive combinations:

\begin{equation}
\mathbf{X}_{j} = \sum_{i=1}^r \mathbf{H}_{ji} \mathbf{W}_{i}, \quad \forall \mathbf{H}_{ji} \geq 0
\end{equation}
where, $\mathbf{X}_{j}$ is the j-th column of matrix $\mathbf{X}$, and $\mathbf{W}_{i}$ is the i-th column of matrix $\mathbf{W}$.

NMF decomposes the data into components that are additive and non-negative, which is well-suited to fMRI data. Signals in fMRI data are usually additive (i.e., the total signal is a sum of signals from different sources). Furthermore, components identified by NMF can be directly mapped to specific brain regions or networks, providing a clear and interpretable characterization of brain activity.

\subsection{Overview of proposed framework}
\label{subsec:framework}
\begin{figure}[htp]
    \centering
    \includegraphics[width=0.95\textwidth]{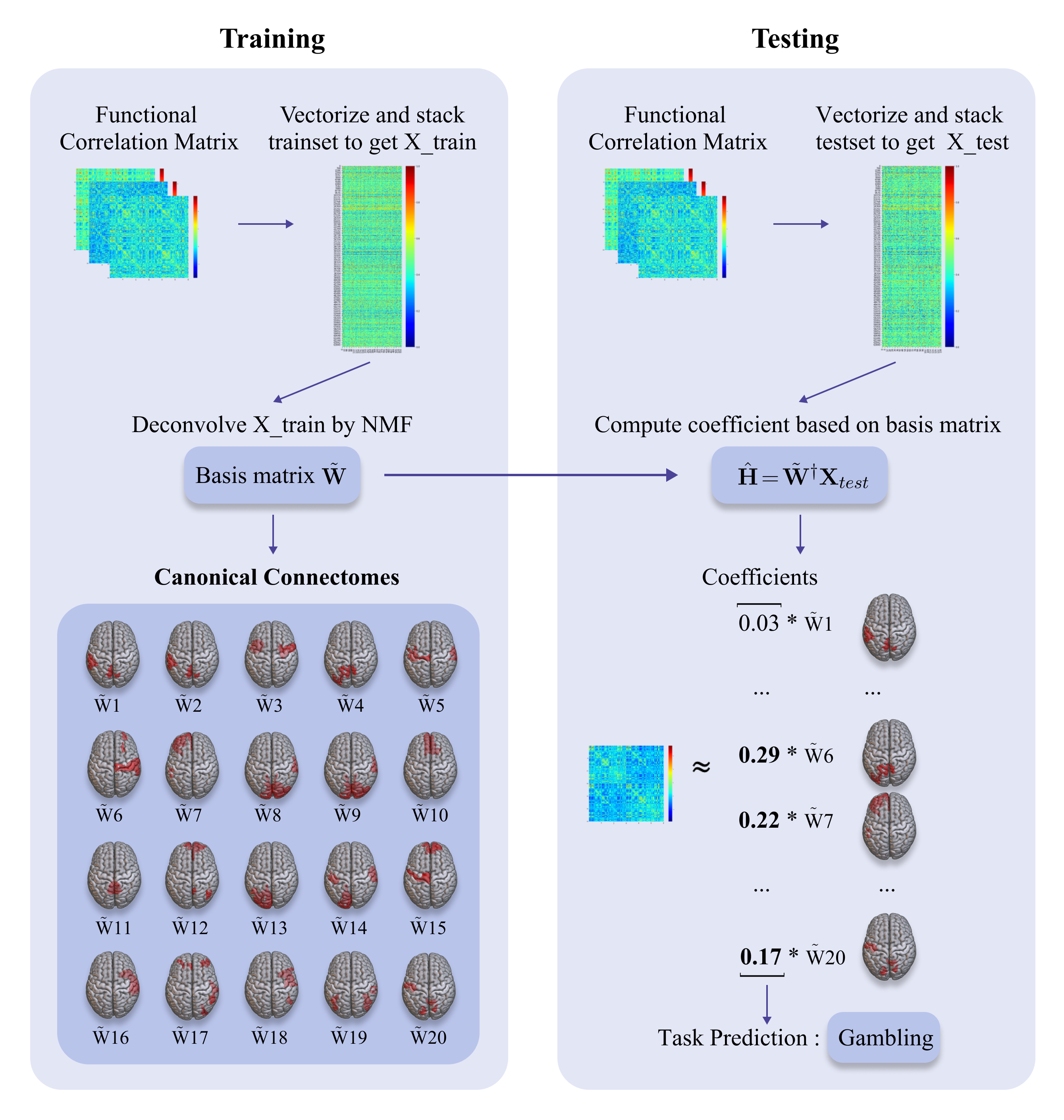}
    \caption{\emph{Overview of proposed framework:} (1) The training phase deconvolves the data matrix of vectorized connectomes in the training set into a small number of basis vectors; (2) The testing phase computes the coefficients of the functional basis and predicts the task on new subjects.}
    \label{fig:overview}
\end{figure}

In Fig. \ref{fig:overview}, we illustrate the general framework. 
We vectorize the connectomes and stack them column-wise into a population-level data matrix $\mathbf{X} \in \mathbb{R}^{\mathcal{O}(d^2) \times n}$, $d$ denotes the number of regions in each connectome, $n$ denotes the total number of data samples.

We decompose the matrix $\mathbf{X}$ into two matrices $\mathbf{W}$ and $\mathbf{H}$ using NMF, as discussed in Section \ref{subsec:nmf}. The columns of the basis matrix $\mathbf{W}$ represent canonical connectomes that can be linearly combined to approximate the original functional connectomes in $\mathbf{X}$. Columns of matrix $\mathbf{W}$ correspond to "canonical task connectomes" since they capture fundamental networks across various subjects.

To validate the generalizability of these canonical representations, we divide the cohort randomly into train and test sets. We first use training data to compute the canonical task connectomes via NMF. These connectomes serve as an interpretable basis to represent brain activity. For the test set, we find the coefficients that best reconstruct each test sample using the pre-computed connectome basis. We then train a model to map from coefficient patterns to cognitive task labels using the training data. Finally, we apply this model and use the resulting coefficients to predict the cognitive tasks performed by subjects in the test set.

Our results show that canonical task connectomes strongly correlate with anatomical and physiological processes associated with different cognitive tasks. The ability to accurately predict tasks in new test data demonstrates that the canonical connectomes generalize across subjects and capture robust task-specific functional networks.

\subsection{Data}
\label{subsec:data}
We validate our model and methods using data from 1000 subjects' fMRIs from the Human Connectome Project (HCP) Young Adult dataset \cite{van13}. For each subject, we have separate fMRIs when they are at rest, and while they perform six cognitive tasks (Language, Emotional Processing, Gambling, Motor, Relational Processing, and Social Processing) \cite{barch13}. We first use the Minimal Pre-Processing Pipeline prescribed by the HCP consortium \cite{glasser13}. This process includes spatial artifact/distortion removal, head motion correction, registration, and normalization to standard space. For each input noisy fMRI, the Minimal Preprocessing Pipeline outputs a clean and standardized $voxel \times time$ time-series matrix. Then, we use the Atlas of Glasser et al. \cite{glasser16} to aggregate this matrix into a $region \times time$ matrix. We note that each \emph{region} consists of proximal \emph{voxel}s with shared anatomy. In all, the Glasser Atlas demarcates 180 regions in each hemisphere (360 in total). We then create the functional connectome (FC) matrix for each fMRI by computing the Pearson Correlation between all pairs of regions. In all, we have 7000 FCs ($1000\ subjects \times 7\ tasks$). We vectorize the upper triangular matrix of each FC and stack them side by side to create a population-level matrix of size $64620 \times 7000$.


\section{Results}
\label{sec:result}

In this section, we show that our canonical task connectomes are conserved across populations, highly specific to a small subset of tasks, and as a consequence provide both an understanding of the neural response, as well as the ability to predict tasks. We provide evidence of strong spatial localization for these representative brain networks, which establishes interpretability on the basis of neuro-anatomy. We also show that the regions implicated in the tasks are supported by prior experimental studies, which establishes physiological interpretations.

\subsection{Canonical Task Connectomes are Generalizable across Cohorts}
\label{sec: Generalizable across Cohorts}
We show that canonical task connectomes are stable representations of different tasks. To demonstrate this, we first compute the canonical representations on a training set. We then predict the task performed in the test set. Specifically, we create $\mathbf{X}_{train}$ and $\mathbf{X}_{test} $ by 80/20 random splits of the subjects. We decompose $\mathbf{X}_{train}$ to find the canonical task connectomes $\tilde{\mathbf{W}}$ and the coefficient matrix $\tilde{\mathbf{H}}$, and use $\tilde{\mathbf{H}}$ along with corresponding task labels to train a classifier. Now, given a test subject (or test set), we compute the least-squares solution $\hat{\mathbf{H}}$, the coefficients corresponding to the test subjects, using $\hat{\mathbf{H}} = \tilde{\mathbf{W}}^{\dagger} \mathbf{X}_{test}$, where $\tilde{\mathbf{W}}^{\dagger}$ is the pseudo-inverse of the matrix of canonical task connectomes from the test set. Finally, we predict the labels of $\mathbf{X}_{test}$ using $\hat{\mathbf{H}}$ and the trained classifier.

\begin{figure}[tph]
    \centering
    \includegraphics[width=0.75\textwidth]{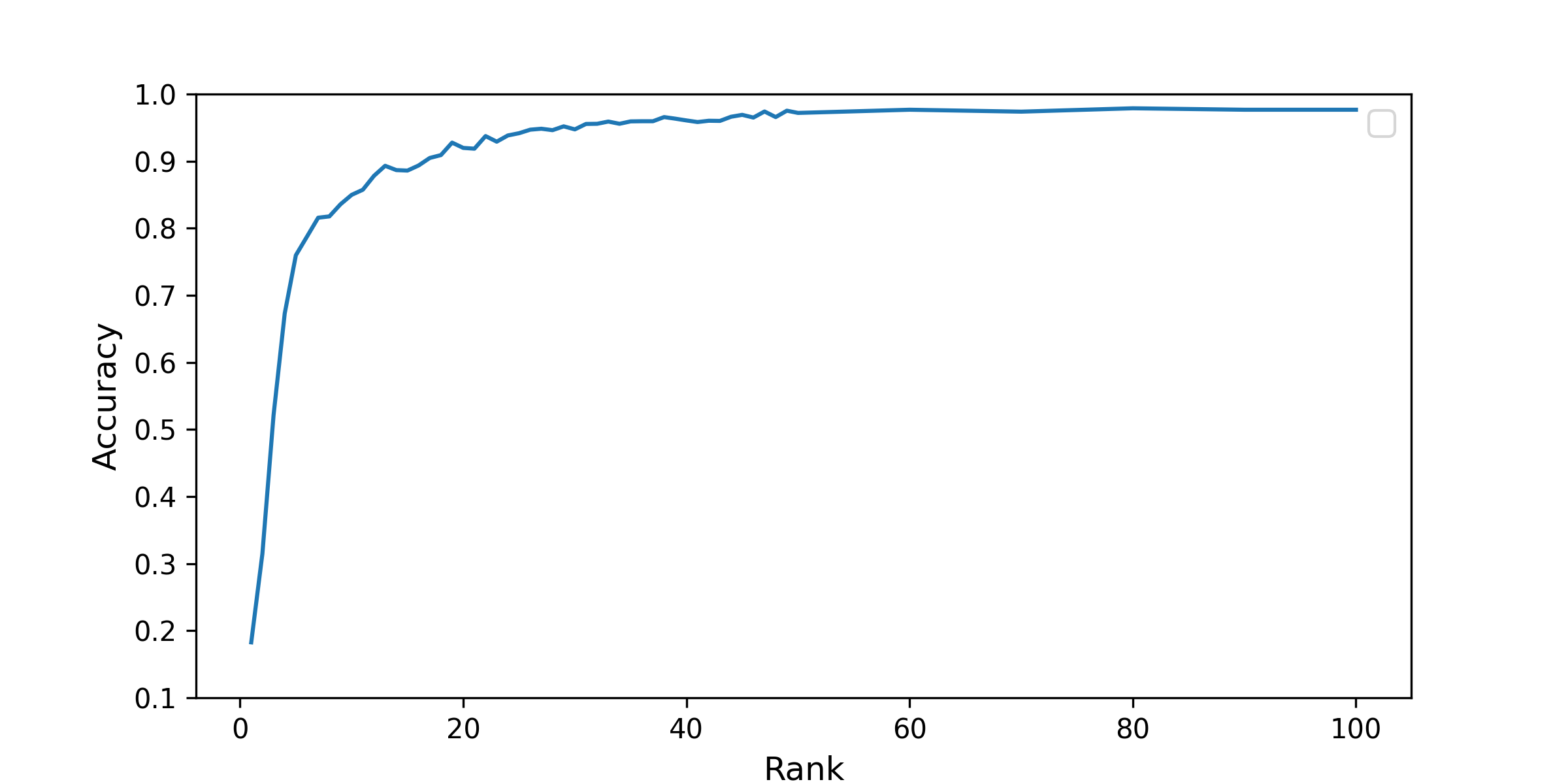}
    \caption{\emph{Parameter study: effect of rank on test accuracy.} We set rank from 1 to 100 and use NMF to deconvolve the train data matrix. We then test task predictions based on $\hat{\mathbf{H}}$ using $\hat{\mathbf{H}} = \tilde{\mathbf{W}}^{\dagger} \mathbf{X}_{test}$. There is rapid increase in accuracy from rank 1 up to 20, after which the rate of increase slows considerably. The accuracy plateaus around 98\% at rank 50.}
    \label{fig:rank_acc}
\end{figure}

\noindent\textbf{Parameter study of rank:} To determine the optimal rank, we decompose the data matrix using NMF with rank ranging from 1 to 100. We then test the accuracy of predicting the task labels from the factorized matrices. Figure \ref{fig:rank_acc} illustrates the prediction accuracy with increasing rank. As shown, the accuracy increases rapidly from rank=1 to 20, increasing from approximately 18\% to 92\%. Subsequently, it exhibits a gradual rise, reaching approximately 98\% once the rank surpasses 50. For our main results in this paper, we select rank=20 as it marks an inflection point and facilitates interpretation, without overfitting to data.

We demonstrate the effectiveness of our method for task prediction and feature selection using three simple classifiers: K-nearest neighbor (KNN), support vector machine (SVM), and a 3-layer multilayer perceptron (MLP). Our approach achieves excellent performance on task prediction and is able to identify the most relevant features for each task. 

In Table \ref{tab}, the first row summarizes the results for rank-20 approximations, averaged over 10 runs. The factors computed by NMF yield high prediction test accuracy by all three classifiers. In the following rows, we demonstrate the strong feature selection ability of our method. We sort and eliminate the 90\%, 95\% and 99\% smallest values in the 64620 feature dimensions of $\mathbf{W}_{i}$, for $i \in \{1,...,20\}$.Then, we combine the remaining dimensions of each $\mathbf{W}{i}$ to obtain a region-of-interest matrix with only
$\sim$52000, $\sim$37000, and $\sim$10500 features, reducing $\sim$19.5\%, $\sim$42.7\% and $\sim$83.75\% of the original data matrix. We use significantly reduced set of features selected by our method for task prediction by the same process above. Remarkably, results consistently outperform the original feature set, yielding superior accuracy. We observe that a cut-off of 95\% - 99\% is ideal. To further demonstrate task-specific feature selection, we perform binary classification on the matrix deconvolved only on the specific task matrix $64620 \times 1000$ rather than the entire $64620 \times 7000$ data matrix. We also sort and eliminate 90\%, 95\%, and 99\% of the smallest values in $\mathbf{W}_{i}$, for $i \in \{1,...,20\}$, and use the remaining non-zero dimensions for binary task prediction, with the target task label = 1 and other 6 tasks’ labels represented as 0. Table \ref{tab2} shows the result of binary classification using a 3-layer multilayer perceptron, yielding excellent accuracy $\sim$98\% with our feature selection. 

\begin{table}
    \renewcommand{\arraystretch}{1.2}
    \caption{\emph{Task prediction test accuracy using different classifiers.} }
    \label{tab}
    \centering
    \begin{tabular}{lllll}
    \toprule
        &KNN      &MLP      &SVM      \\  
    \midrule
    Original   & 86.41 $\pm$ 1.05   & 92.31 $\pm$ 0.65  &91.39 $\pm$ 1.18 \\
    Cut 90   & 86.41 $\pm$ 1.41   & 92.26 $\pm$ 0.97  &91.99 $\pm$ 1.01 \\ 
    Cut 95   & 87.24 $\pm$ 1.59   & \textbf{93.16 $\pm$ 0.50}  &\textbf{92.80 $\pm$ 0.94}\\
    Cut 99   &\textbf{87.77 $\pm$ 1.31}  & 93.13 $\pm$ 0.81  &92.39 $\pm$ 0.79 \\
    \bottomrule
  \end{tabular}
\end{table}

\begin{table}
  \centering
  \renewcommand{\arraystretch}{1.2}
  \caption{\emph{Binary classification test accuracy using MLP}}
  \label{tab2}
  \begin{tabular}{lp{1.8cm}p{1.8cm}p{1.8cm}p{1.8cm}p{1.8cm}p{1.8cm}p{1.8cm}}
    \toprule
    &Rest &Language &Emotion &Gambling &Motor &Relational &Social \\  
    \midrule
    Original & 97.30$\pm$0.40 & 98.25$\pm$1.01 & 97.49$\pm$1.41 & 96.83$\pm$1.72 & 96.92$\pm$1.59 & 96.98$\pm$1.45 & 97.36$\pm$1.64 \\
    Cut 90   & \textbf{97.60$\pm$0.31} & \textbf{98.43$\pm$0.87} & \textbf{97.61$\pm$1.45} & 96.96$\pm$1.71 & 97.06$\pm$1.56 & 97.11$\pm$1.44 & 97.47$\pm$1.61 \\ 
    Cut 95   & 97.34$\pm$0.49 & 98.42$\pm$1.14 & 97.57$\pm$1.60 & \textbf{97.14$\pm$1.61} & \textbf{97.12$\pm$1.45} & \textbf{97.25$\pm$1.36} & \textbf{97.61$\pm$1.53} \\
    Cut 99   & 96.10$\pm$0.35 & 97.52$\pm$1.46 & 96.95$\pm$1.48 & 96.52$\pm$1.50 & 96.42$\pm$1.40 & 96.67$\pm$1.41 & 97.12$\pm$1.69 \\
    \bottomrule
  \end{tabular}
\end{table}

\subsection{Canonical Task Connectomes are Building Blocks of Cognitive Tasks}
\label{sec: Building Blocks of Tasks}

We show that canonical task connectomes constitute fundamental building blocks of human cognitive tasks. In the first set of results, we demonstrate that our canonical connectomes encode information that constitutes the observed composite connectome in a manner that is conserved across subjects. This is non-trivial because of: a) inherent heterogeneity in basal brain activity across individuals; b) individual-level differences in cognitive processes to perform a task; c) diversity of task conditions; and d) noise in the imaging modality. 
Using NMF, we deconvolve the population-level matrix $\mathbf{X}$. As mentioned in Section \ref{subsec:framework}, NMF outputs a) a small set of vectorized canonical connectomes $\mathbf{W}$, and b) a linear coefficients matrix $\mathbf{H}$. Each column of $\mathbf{H}_{(\ast,j)}$ represents the extent to which the canonical connectomes contribute to the corresponding vectorized connectome in $\mathbf{X}_{(\ast,j)}$. Additionally, we note that each row $\mathbf{H}_{(i,\ast)}$ represents the linear coefficients for the $i-$th canonical connectome, i.e. $\mathbf{W}_{(\ast,i)}$.
In our experiments, we set the rank to 20 to extract a compact set of informative task networks. 

\begin{figure}[tph]
    \centering
    \includegraphics[width=0.9\textwidth]{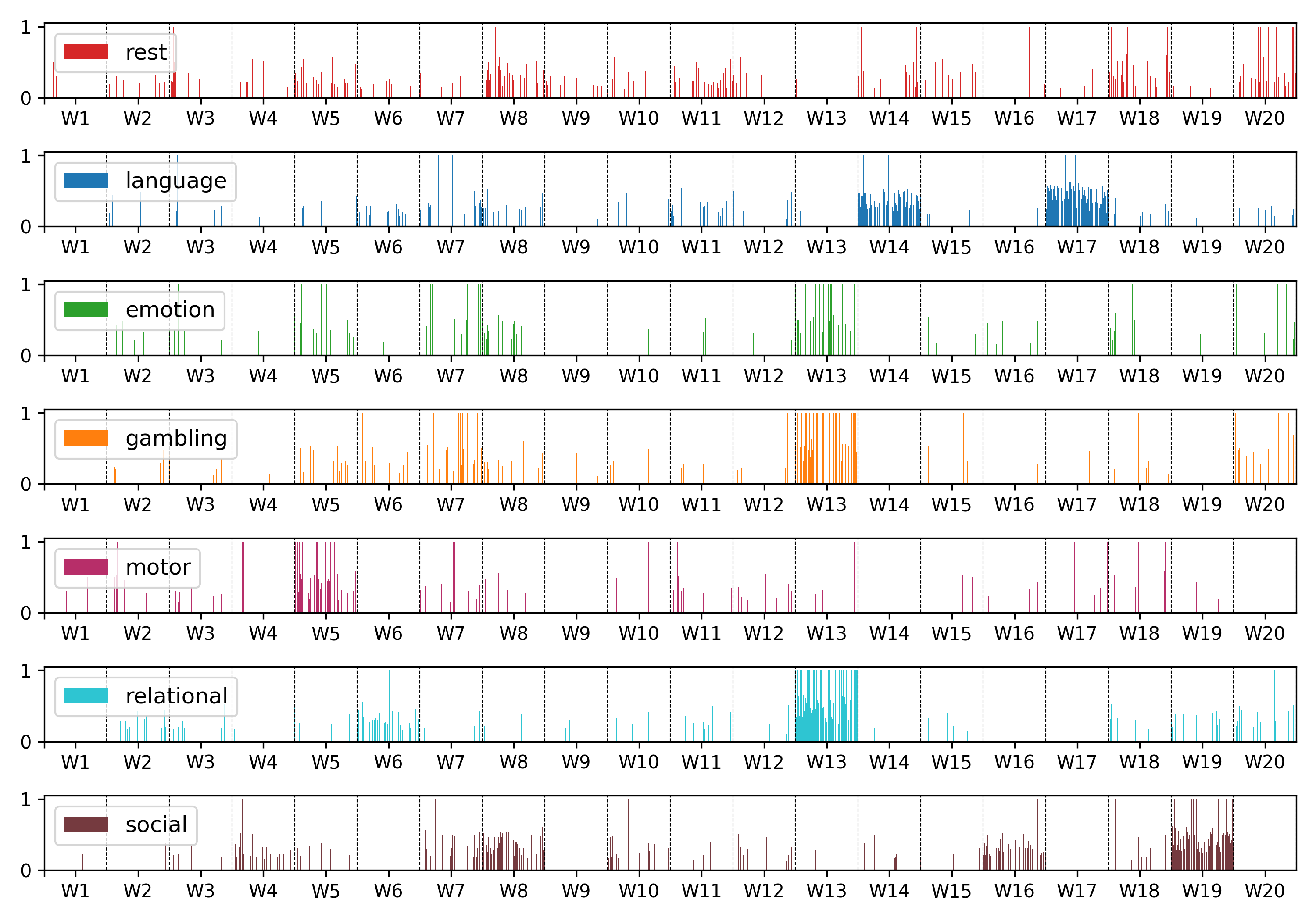}
    \caption{\emph{Coefficients matrix $\mathbf{H}$ of NMF.} 
    We apply NMF to factor the data matrix of rank 20. We fit columns of $\mathbf{H}$ to a normal distribution and use the 90th percentile as a cutoff to discard small values. Each plot shows the encoding of 1000 subjects performing one task, visualizing $\mathbf{W}_i$ (corresponding to the canonical connectomes) that make up that particular task.}
    \label{fig:nmf_90percentile}
\end{figure}

We z-score normalize the columns of $\mathbf{H}$ returned by NMF 
and retain only those non-zero values higher than 90 percentile.  
The rows of $\mathbf{H}$, grouped according to tasks are visualized in Fig. \ref{fig:nmf_90percentile}. 
It is evident that the non-zeros of these significant coefficients are highly selective of tasks. We use the 90th percentile as a cutoff to discard small values and select the most significant coefficients. This allows us to identify the canonical connectomes most strongly associated with each task (at the 90th percentile significance level) which can be seen in Table \ref{table:significant_connectomes}.

\begin{table}[hpt]
\caption{\emph{Significant Connectomes for Each Cognitive Task}}
\renewcommand{\arraystretch}{1.2}
\centering
\begin{tabular}{ll}
\toprule
\textbf{Task}    & \textbf{Significant Connectomes}          \\ 
\midrule
Rest             & W3, W8, W9, W11, W15, W18, W20            \\ 
Language         & W14, W17                                  \\ 
Emotion          & W7, W13, W20                              \\ 
Gambling         & W6, W7, W13, W20                          \\ 
Motion           & W1, W2, W4, W5, W11, W12, W18             \\ 
Relational       & W6, W13                                   \\
Social           & W4, W8, W10, W16, W19                     \\ 
\bottomrule
\end{tabular}
\label{table:significant_connectomes}
\end{table}


\begin{figure}[tph]
    \centering
    \includegraphics[width=0.85\textwidth]{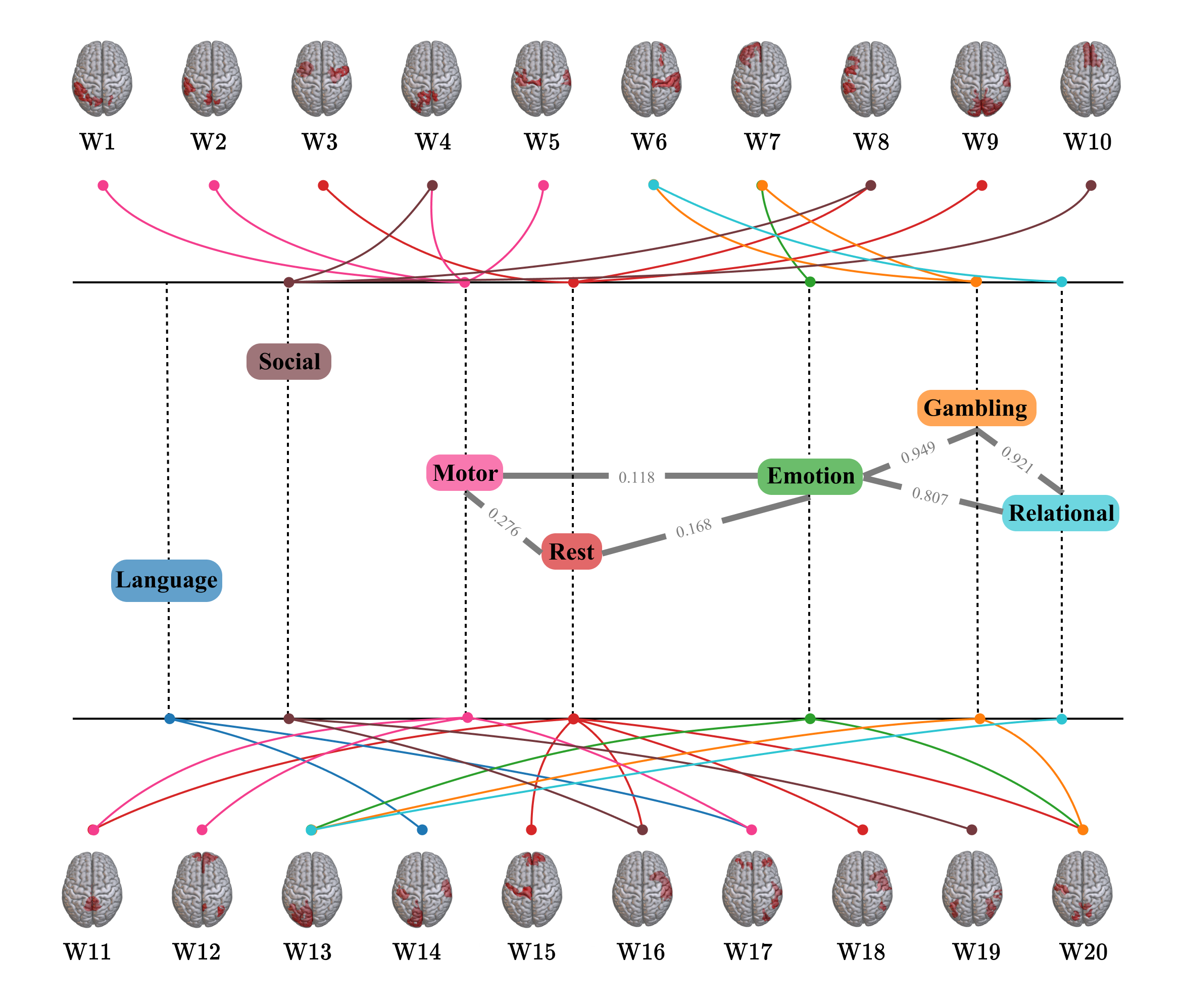}
    \caption{\emph{Correlation network analysis.} We take the mean of each task's corresponding linear coefficients and compute the Pearson coefficient across these averaged task vectors. We then construct the similarity network using the Pearson coefficient and connect each task with its most significant canonical connectomes. }
    \label{fig:graph}
\end{figure}

We compute the similarity of fMRI responses for different tasks by constructing a network in which nodes correspond to tasks and edges correspond to similarities computed using our coefficient vectors. Specifically, if the coefficient vectors for two tasks are similar, they are composed of the same set of canonical connectomes in similar proportion. To demonstrate this, we begin by averaging each of the seven cognitive tasks' coefficient vectors across 1000 subjects, obtaining a set of task-specific vectors from $\tilde{\mathbf{H}}$. We then calculate the Pearson correlation coefficients between these vectors and construct the task similarity network. 
%
Fig. \ref{fig:graph} shows the task similarity network along with the canonical connectomes associated with each task. 
%
This network reveals high similarity between Emotion, Relational, and Gambling tasks. Rest, Emotion and Motor are somewhat related, while Social and Language tasks exhibit distinct characteristics from the others, indicating distinct brain network engagements for these functions.

\subsection{Canonical Task Connectomes have a Strong Anatomical and Physiological Basis}
\label{sec: Anatomical and Physiological Basis}
We show that: a) each canonical task connectome is spatially localized to anatomically demarcated regions; and b) the regions enriched in each canonical connectome are known to be implicated in the corresponding task. As before, we deconvolve the population-level matrix $\mathbf{X}$ to compute $\mathbf{W}$ and $\mathbf{H}$. In this experiment, we use rank 20 approximation for interpretation. From each column $\mathbf{W}_{i}$, we construct $region \times region$ canonical task connectome, which is a correlation matrix and we sum the rows to compute the active factor for each region and retain the top 2\% of them. We then use MRIcroGL~\cite{rorden2000stereotaxic} for visualization of the top active region for each task.

\begin{figure}
    \centering
    \includegraphics[scale=0.8]{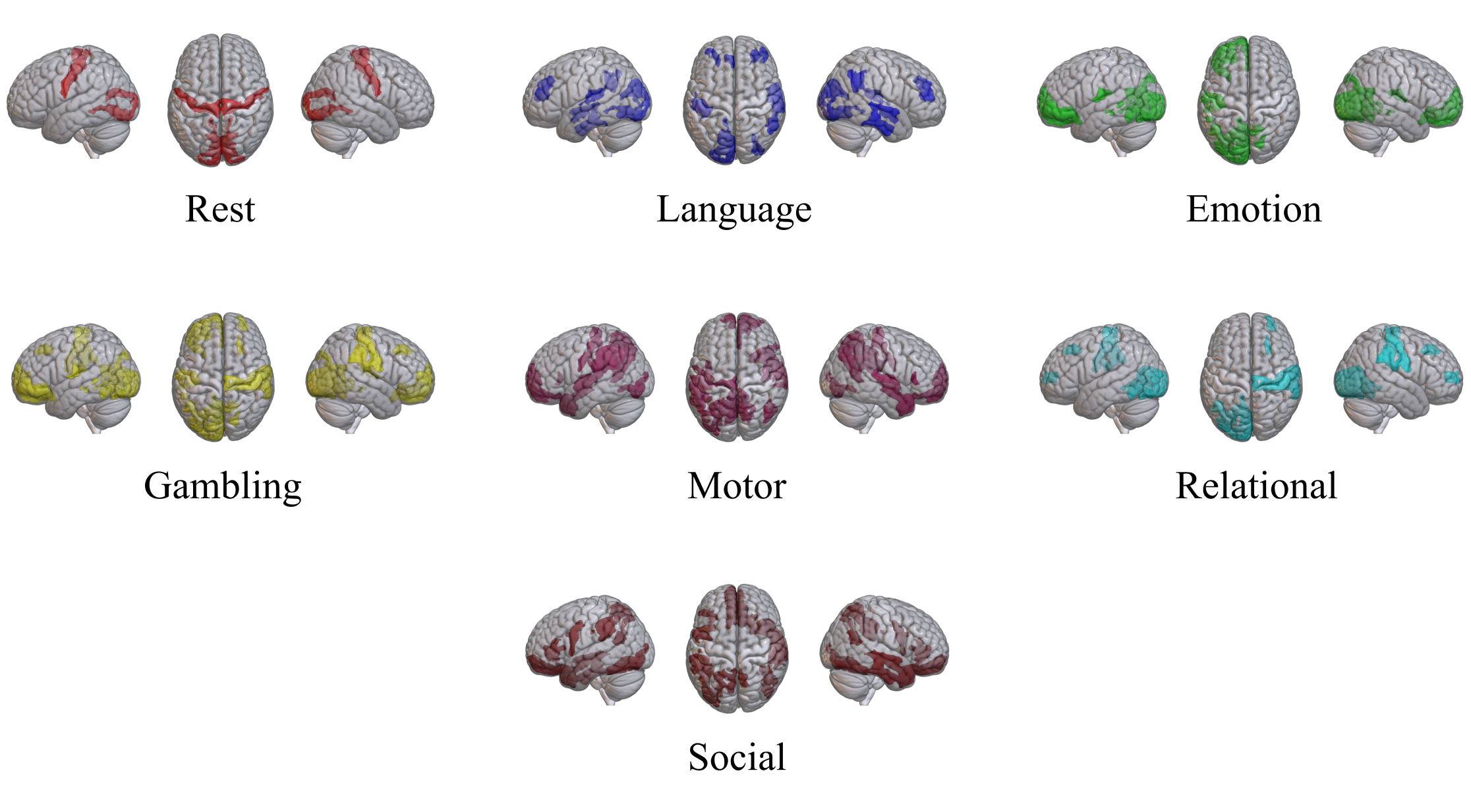}
    \caption{\emph{Canonical Task Connectomes have strong anatomical basis.} From each column $\mathbf{W}_{i}$, we construct $region \times region$ canonical task connectome. The correlation values for the 360 regions are summed as an active factor, and the top 2\% of highly active regions are retained. MRIcroGL is then used to visualize the most active regions for each task. }
    \label{fig:brain_region}
\end{figure}

In Fig \ref{fig:brain_region}, we visualize the task-specific connectomes by combining canonical task connectomes that we find significant for each task in Section \ref{sec: Building Blocks of Tasks}.

Our results indicate that anatomical regions indicative of rest correspond to the dorsomedial prefrontal cortex, posterior cingulate cortex (in the limbic node), the angular gyrus found in the posterior part of the inferior parietal lobe, and substructures corresponding to the dorsal medial prefrontal cortex. These regions are also implicated in the dorsal Default Mode Network (dDMN) -- a functional network known to be active during Rest \cite{biswal95}. 
In the language processing task, our approach identifies regions in the inferior parietal cortex/angular gyrus, which are well known to be active during language processing tasks \cite{seghier13}. Further, regions in the left prefrontal cortex are associated with word and sentence comprehension \cite{gabrieli98}. We also observed regions in the auditory cortex and auditory association cortex which explains the response to auditory stories. 
The regions implicated for emotion processing in our setup include anterior cingulate and ventromedial prefrontal cortex, which are both cortical regions known to be active when processing facial expressions. Subcortical regions such as Amygdala, also commonly reported to be active during emotional processing are not considered as we limit our study to the cortex \cite{vsimic21}.
Since gambling tasks are cognitively complex, several brain regions are co-activated. Our results demonstrate that subregions within posterior and anterior cingulate, orbital frontal cortex and dorsolateral prefrontal cortex are all implicated in gambling related tasks, which is in line with observations of Li et al. \cite{li2010iowa}.
The regions implicated in relational processing are dorsolateral Prefrontal Cortex rostrolateral prefrontal cortex and posterior parietal cortex, which are consistent with previous literature \cite{holyoak21}. 
The regions implicated in social processing are the medial prefrontal cortex, which is located in the prefrontal cortex of the frontal lobe \cite{frith07}, fusiform gyrus, and anterior cingulate cortex \cite{tso18}.

\section{Related Work}
\label{sec:related}
\vspace{-3mm}
\textbf{Matrix Factorization.}
Independent Component Analysis (ICA) and its variants are widely used in fMRI analysis. Spatial Independent Component Analysis (ICA) \cite{makeig95,Zhang2019ExperimentalCO, Smitha2019RestingFA, Calhoun2012MultisubjectIC, Calhoun2006UnmixingFW} methods decompose fMRI data into a set of spatially independent components. They identify patterns of activity across the brain that are independent of one another. This information is used to identify distinct networks of brain regions involved in various cognitive processes.
In a typical ICA model, the source signals are assumed to be statistically independent and non-Gaussian, with an unknown linear mixing process. The model assumes that every observed vector $x \in \mathbb{R}^m$ is generated by a linear mixture of $n$ independent sources $x = \mathbf{A}s$, where 
$s \in \mathbb{R}^n$ is an N-dimensional vector whose elements are the random variables that refer to the independent sources and $\mathbf{A} \in \mathbb{R}^{m \times n}$ is an unknown mixing matrix. ICA aims to estimate an unmixing matrix $\mathbf{W} \in \mathbb{R}^{n \times m}$ such that the recovered sources: $y= \mathbf{W}x = \mathbf{W A}s$ is a good representation of the true sources $s$. 
Applying the typical ICA model to fMRI data, we have data $\mathbf{X} = \mathbf{AS}$, where $\mathbf{X} \in \mathbb{R}^{N \times V}$ spans $N$ time points and $V$ voxels, and $\mathbf{S}$ contains spatially independent source signals.

Group ICA is an extension of spatial ICA that allows the identification of common patterns of activity across multiple subjects in a study. A popular implementation of Group ICA is Multivariate Exploratory Linear Decomposition into Independent Components (MELODIC) \cite{beckmann04}, which is part of the fMRI Standard Library (FSL).
Other approaches for multi-subject analysis using ICA have been proposed \cite{Calhoun2001AMF, Esposito2005IndependentCA, Guo2008AUF, Schmithorst2004ComparisonOT, Beckmann2005TensorialEO}. 
The model in Calhoun et al. \cite{Calhoun2001AMF} defines Group ICA as $\mathbf{X}_i = \mathbf{A}_i \mathbf{S} $, where $\mathbf{X}_i \in \mathbb{R}^{N_i \times V}$ is the fMRI observation for subject $i$ with $N_i$ time points and $V$ voxels. Group ICA captures a group subspace with independent spatial maps and time courses. Then, these are used to reconstruct subject-specific spatial maps $\mathbf{S}_i$ and time courses $\mathbf{A}_i$. 
Group ICA has been widely used to study functional connectivity differences between groups of healthy and clinical populations \cite{To2021TowardsDG, Du2013GroupIG}, as well as to identify brain networks associated with specific cognitive processes across a group of individuals \cite{Elseoud2013ExploringFB, Jung2021DistinctBC}. 
We rely on NMF in this study instead of more commonly used ICA and related methods, because NMF requires that all components in the decomposition be strictly positive, which along with sparsity of factors yields interpretability. 

\noindent\textbf{Deep learning methods.}
Deep learning models like MLPs and CNNs have been used to extract high-level fMRI features for classification ~\cite{suk2016state,huang2017modeling}. More commonly, classifiers use functional connectivity features between brain ROIs to improve performance ~\cite{meszlenyi2017resting,xing2019dynamic}. Graph neural networks, with ROIs as nodes and FC as edge weights, are gaining popularity for their alignment with the brain's structure ~\cite{li2021braingnn,li2019graph}. Recurrent and transformer models have also been added to capture temporal dynamics in functional connectivity ~\cite{kim2021learning}.
However, with the increasing size of parameters, interpretability of these networks remains an unresolved issue. Our framework explains observed (composite) brain activity in terms of canonical task connectomes, which have biological basis -- a key contribution of our work.

\noindent\textbf{Other methods.}
Subspace clustering methods are used in fMRI to partition data into subspaces and assign each data point (e.g., voxel or region of interest) to its corresponding subspace. This allows for the identification of different brain activity patterns or functional connectivity profiles within data. Several subspace clustering methods have been applied to fMRI data such as spectral clustering \cite{Gupta2020IterativeCS, Craddock2012AWB,Alsharoa2018TemporalBS}, sparse subspace clustering \cite{Sui2016LocalityRS, Kim2017JointFA}, low-rank and sparse decomposition (LRSD) \cite{Tu2022LowRankPS, Uruuela2021ALR, Singh2015UndersampledFM}.
Subspace clustering reveals distinct brain activity patterns, functional networks, or connectivity profiles within the data. In our study, NMF is preferable to subspace clustering for fMRI analysis due to its simplicity, interpretability, and computational efficiency. The nonnegative parts-based decomposition from NMF provides intuitive and meaningful canonical task connectomes that subspace clustering methods cannot.

\vspace{-3mm}

\section{Conclusion}
\label{sec:conclusion}
\vspace{-3mm}
We presented a new problem and framework for fMRI analysis that deconvolves an input set of neuroimages of subjects performing different cognitive tasks into a compact set of task-specific elementary networks called ``canonical task connectomes''. Our approach formulates the problem as one of matrix factorization, revealing that the resulting latent factors/networks can serve as fundamental ``building blocks'' for various cognitive tasks. Experimental results conducted on the Human Connectome Project dataset demonstrate how NMF captures the inherent task-specific structure in suitably abstracted neuroimages. Furthermore, we illustrate the utility of these canonical task connectomes as effective biomarkers for predicting the performed task. Additionally, we show anatomical and physiological underpinnings for the networks identified by our framework.
Our framework can be extended to more complex applications, such as: a) understanding shared and unique functional networks across different pathologies; and b) how task-specific networks can get dysregulated due to the onset, and progression of diseases.

\bibliographystyle{unsrt}  
\bibliography{main}

\end{document}